\documentclass[a4paper]{article}

\usepackage{INTERSPEECH2020}
\usepackage{url}
\usepackage{amsmath,graphicx}
\usepackage{dsfont}
\usepackage{multirow}
\usepackage{arydshln}
\usepackage{booktabs}
\usepackage{hyperref}
\usepackage{enumitem}
\usepackage[tracking=true]{microtype}
\usepackage[table,xcdraw]{xcolor}
\usepackage{boldline}
\usepackage{enumitem}
\usepackage{setspace}

\newcommand\narrowstyle{\SetTracking{encoding=*}{-30}\lsstyle}

\title{A Differentiable Perceptual Audio Metric \\ Learned from Just Noticeable Differences}
\name{\narrowstyle Pranay Manocha$^{1}$ \hspace{0mm} Adam Finkelstein$^{1}$ Richard Zhang$^{2}$ \hspace{0mm} Nicholas J. Bryan$^{2}$ \hspace{0mm} Gautham J. Mysore$^{2}$ Zeyu Jin$^{2}$ \hspace{0mm}}
\address{$^{1}$Princeton University \hspace{30mm}$^{2}$Adobe Research}
\vspace{-0.25\baselineskip}
\email{~~~~~~~~~~~$^{1}$\{pmanocha,af\}@cs.princeton.edu ~~ $^{2}$\{rizhang,nibryan,gmysore,zejin\}@adobe.com}

\newcommand{\ignorethis } [1] {}

\newcommand{\etal       }     {{et~al.}}

\newcommand{\eg         }     {{e.g.}}

\newcommand{\Reals      }     {{\textrm{I\kern-0.18em R}}}

\renewcommand{\vec      } [1] {{\text{\boldmath $\mathbit{#1}$}}}

\newcommand{\change     } [1] {\mbox{{\footnotesize $\Delta$} \kern-3pt}#1}

\newlength{\w}

\newcommand{\colornote}[3]{{\color{#1}\bf{#2: #3}\normalfont}}
\newcommand {\Pranay}[1]{\colornote{violet}{P}{#1}}

\newcommand{\beginsupplement}{%
        \setcounter{table}{0}
        \renewcommand{\thetable}{S\arabic{table}}%
        \setcounter{figure}{0}
        \renewcommand{\thefigure}{S\arabic{figure}}%
     }

\begin{document}

\maketitle
\begin{abstract}
Many audio processing tasks require perceptual assessment. The ``gold standard'' of obtaining human judgments is time-consuming, expensive, and cannot be used as an optimization criterion. On the other hand, automated metrics are efficient to compute but often correlate poorly with human judgment, particularly for audio differences at the threshold of human detection. In this work, we construct a metric by fitting a deep neural network to a new large dataset of crowdsourced human judgments. Subjects are prompted to answer a straightforward, objective question: are two recordings identical or not? These pairs are algorithmically generated under a variety of perturbations, including noise, reverb, and compression artifacts; the perturbation space is probed with the goal of efficiently identifying the just-noticeable difference (JND) level of the subject. We show that the resulting learned metric is well-calibrated with human judgments, outperforming baseline methods. Since it is a deep network, the metric is differentiable, making it suitable as a loss function for other tasks. Thus, simply replacing an existing loss (e.g., deep feature loss) with our metric yields significant improvement in a denoising network, as measured by subjective pairwise comparison.

\end{abstract}
\iffalse
\fi
%\vspace{-0.1in}
\section{Introduction}
\label{sec:intro}
\begin{figure}[h!]
\vspace{-0.5\baselineskip}
\centering
\setlength{\w}{0.32\columnwidth}
\setlength{\tabcolsep}{2pt}
\begin{tabular}{ccc}
\includegraphics[width=\w]{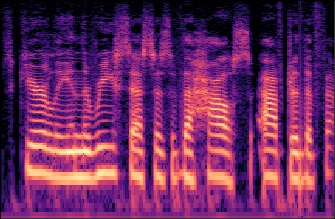} &
\includegraphics[width=\w]{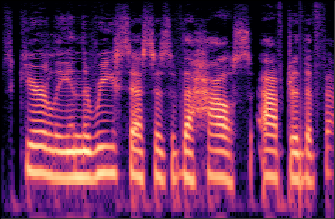} &
\includegraphics[width=\w]{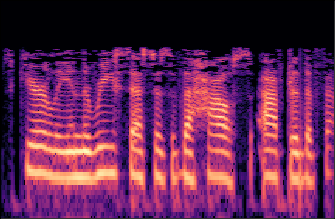} \\
\begin{minipage}{\w}
\vspace{2pt}
\begin{spacing}{0.9}
{\footnotesize {\bf Audio1} closer to Ref. according to {\bf humans}, deep embeddings, and {\bf our model}.}
\end{spacing}
\end{minipage} &
\begin{minipage}{\w}
\centering
{\footnotesize {\bf Reference} \\ speech recording. \\ ~ \\ ~ } 
\end{minipage} &
\begin{minipage}{\w}
\vspace{2pt}
\begin{spacing}{0.9}
{\footnotesize {\bf Audio2} closer to Ref. according to L1, L2,
PESQ \& ViSQOL. \\ ~ }
\end{spacing}
\end{minipage}
\end{tabular}
\vspace{-3.5ex}
\caption{
Audio1 contains white noise, whereas Audio2 is band-limited: 
which one sounds ``closer'' to the reference recording? 
Existing objective metrics (\eg, L1, L2, PESQ and ViSQOL) struggle to measure JNDs, and often disagree with human judgments, unlike deep embeddings and our model.}
\label{model}
\vspace{-4ex}
\end{figure}

Humans have an innate ability to analyze and compare sounds.  While efforts have been made to emulate human judgment via automatic methods, the gap between human and machine judgment remains large [see Figure~\ref{model}]. This gap is acute in the context of synthetic audio based on deep learning~\cite{oord2016wavenet}, which has become so realistic that most metrics fail to reflect human perception. Many deep learning models rely on a metric for their loss functions; and misalignment between the loss and human judgment yields audible artifacts. Thus, the need for a perceptually-consistent metric hinders advancement of audio processing.

Based on human assessment studies, researchers have developed metrics that evaluate sound quality relative to a reference recording, \eg\ PESQ~\cite{rix2001perceptual}, POLQA \cite{beerends2013perceptual} and ViSQOL~\cite{hines2015visqol}. However, these methods suffer from two general drawbacks.  First, these models have acknowledged shortcomings such as sensitivity to perceptually invariant transformations~\cite{hines2013robustness,manjunath2009limitations}, which hinders stability in more diverse tasks such as speech enhancement. Second, these metrics are non-differentiable, and thus cannot be directly leveraged as a training objective within the context of deep learning. 

Addressing the latter concern,
researchers have trained differentiable neural networks that incorporate such perceptual models,
for example estimating PESQ at each training iteration~\cite{zhang2018training,fu2019learning}. 
The approach of Zhang~\etal~\cite{zhang2018training} encumbers training with expensive
gradient estimation at each step, whereas that of Fu~\etal~\cite{fu2019learning} 
fails to model unseen perturbations.

An alternative is to learn a loss function via adversarial learning (GANs), which 
has shown promising results in enhancement~\cite{pascual2017segan}, synthesis~\cite{donahue2018adversarial}, and source separation~\cite{stoller2018adversarial}.
Another approach adapts the \emph{deep feature loss}~\cite{gatys2015neural} notion from the computer vision community, by using representations learned from a different task to construct similarity metrics~\cite{zhang2018unreasonable}. This idea has been adopted for various audio tasks~\cite{germain2018speech,ananthabhotla2019towards}. 
However, these methods are problem-specific~\cite{kilgour2018fr,avila2019intrusive} and require human assessment for accurate evaluation, particularly when small perceptual differences need to be measured.

% P3: Summarization of our contributions to perceptual audio metrics and deep learning

We propose a new perceptual audio metric based on just-noticeable differences (JNDs) -- the minimal change at which a difference is perceived. To do so, we first collect a large scale dataset of human judgments wherein subjects are asked an easy question: whether two audio recordings sound the same or different. Recordings are modified by injecting various perturbations characteristic of degradations commonly found in audio processing tasks, including noise, reverb, equalization distortion, and compression. The data collection process relies on active learning to efficiently sample such artifacts near the JND level. Next, we train a neural network with this data, and use the learned representation to construct a distance metric that measures the difference between two audio signals. We validate the new metric by showing that it correlates well with three diverse existing mean opinion score (MOS) datasets, as well as three two-alternative forced choice test (2AFC) datasets. Finally we show that using the new metric as a loss function improves the performance of a state of the art denoising network.

%that is better aligned with human judgments 

%\vspace{.05in}
%\noindent 
Thus, our contributions are as follows:
%\vspace{-0.05in}
%\begin{enumerate}[leftmargin=0.44cm,noitemsep]
%
(1)~a framework for collecting crowdsourced human JND judgments for audio recordings;
(2)~a differentiable \textit{perceptual loss model} trained on these data;
(3)~experiments showing that this model correlates better with MOS tests than standard metrics;
(4)~demonstrable improvement in a state-of-the-art speech enhancement network wherein the loss function is enhanced by our model; and
(5)~the dataset, code and resulting metric, as well as listening test examples -- are all available from our project page: \\ {\footnotesize\tt\narrowstyle \href{https://pixl.cs.princeton.edu/pubs/Manocha_2020_ADP/}{http://pixl.cs.princeton.edu/pubs/Manocha\_2020\_ADP}
}
%\end{enumerate}

\iffalse
\item Our metric, data and code can be found at \url{https://pixl.cs.princeton.edu/pubs/Manocha_2020_ADP/}
\fi

\iffalse
We show that our denoiser shows non-trivial improvement from state-of-the-art methods via pairwise comparison listening test. Our metric, data and code can be found at \url{https://pixl.cs.princeton.edu/pubs/Manocha_2020_ADP/}.
\fi

%\vspace{-2\baselineskip}
\section{Proposed Framework}
\label{ssec:framework_main}
We collect a dataset of human judgments using  crowdsourcing tools, which have been shown to perform similarly to expert, in-lab tests~\cite{cartwright2016fast, cartwright2018crowdsourced}; and then we fit a model to these data.
\subsection{Data collection via active learning}
\label{ssec:framework1}
\begin{figure}[t!]
\centering
\includegraphics[width=\columnwidth]{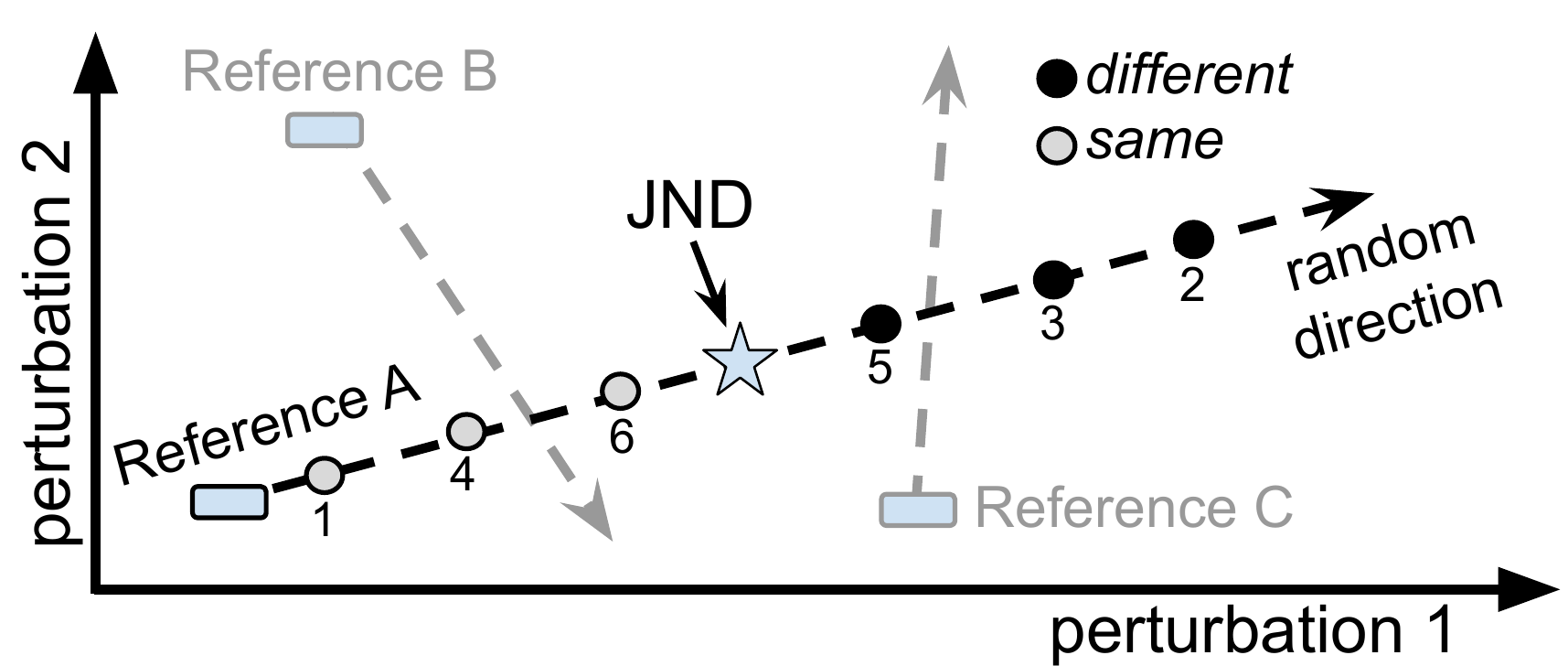}
\vspace{-1.5\baselineskip}
\caption{Depiction of active learning for data collection. From each reference, we probe along a random vector of combined perturbations (dashed line) to search for the JND. Sequential listener responses (numbered) indicate recordings sound the same as (hollow dots) -- or different from (solid) -- the reference.}
\label{fig:activelearning}
\vspace{-\baselineskip}
\end{figure}

We present a listener with two recordings, a reference $x_\mathtt{ref}$ and perturbed signal $x_\mathtt{per}$, and ask if these two recordings are \textit{exactly same or different}, and record the binary response $h\in \{0,1\}$.
For the reference recording $x_\mathtt{ref}$, we first sample a speech recording from a large collection and then degrade it by randomly applying a set of perturbations (e.g. noise and reverb). To produce the perturbed recording $x_\mathtt{per}$, we select a perturbation direction, or ``axis'' which can be one of several perturbation types or a combination applied sequentially. Figure~\ref{fig:activelearning} shows an example where the perturbation direction is a combination of two perturbation types. The types we study are further described in Section~\ref{sec:perts}. The perturbed recording $x_\mathtt{per}$ is produced as a function $\mathcal{H}$ of strength $\rho \in [0,100]$, $x_\mathtt{per}=\mathcal{H}(x_\mathtt{ref}, \rho)$.

For values of $\rho$ that are too large or small, the answer is ``obviously'' different or the same, respectively, and a downstream metric is unlikely to gain information from such data. As such, we employ an active learning strategy to efficiently gather labelled data, in contrast to past approaches~\cite{mcshefferty2015just}. Our goal is to identify the just noticeable difference (JND) threshold, $\rho_{\mathtt{jnd}}$, 
such that a subject can \textit{just} hear the difference between $x_\mathtt{ref}$ and $x_\mathtt{per}$. We attempt to sample $\rho$ to be close to the JND point, illustrated at a high-level in Figure~\ref{fig:activelearning}.

We estimate the current subject's most likely JND $ \rho_{\mathtt{jnd}}^*$, based on all past answers, and then produce the next test case by $x_\mathtt{per}=\mathcal{H}(x_\mathtt{ref}, \rho_{\mathtt{jnd}}^*)$. We assume that human answers follow a Gaussian distribution with mean $\mu$ at the JND point and variance $\sigma^2$, representing human error. 
Following this, we compute the likelihood of $N$ past answers using $\mathcal{L}(\mu,\sigma^2) = \prod_{j=1}^N (1-h_j)(1-c({\rho}_j|\mu,\sigma^2)) + h_j c({\rho}_j|\mu,\sigma^2)$, where ${\rho}_1, ..., {\rho}_N$ are past perturbation strengths,
% used to produce $x_\mathtt{per}$,
$h_1,...,h_N$ are the human judgments, and
$c({v}_j|\mu,\sigma^2)$ is the CDF
% of JND distribution modeled in
of Gaussian $\mathcal{N}(\mu,\sigma^2)$. After computing $\mu$ and $\sigma$ to maximize the above likelihood function,
the next test case follows from ${\rho}_{\mathtt{jnd}}^*=\mu$. The ultimate product of our data collection is a database of triplets $\{x_\mathtt{ref},x_\mathtt{per},h\}$, which we leverage for training a perceptual metric.
\iffalse
Additionally, we put priors on $\mu$ and $\sigma$ to make the first several tests less susceptible to human error. This has several advantages:
\begin{itemize}[leftmargin=0.33cm]
  \item \textbf{Information Maximization}: one good way to achieve maximum information gain is to ask questions around JND, where the answers are the least obvious and most challenging. This strategy has the advantage of inherently creating a ``balanced`` dataset~\cite{roy2001toward} where you have an almost equal number of ``same`` or ``different`` answers.
  \item \textbf{Extra added bias}: We also encourage an equal chance of saying \textit{same} or \textit{different} by using $v_{\mathtt{jnd}}^*=\mu+q\sigma$, where $q > 0$ when we have collected more ``same'' than ``different'', and vice versa. This is done so that the participant is likely to break the trend of giving same answers. If the same trend still continues, we discard that participants' data, as they are not paying attention.
  
  \item \textbf{Additional Priors}: Our model also starts out with a prior that focuses on exploration early on in the test. As more data is acquired, the model becomes more confident and the prior is deemphasized. This procedure (a) stochastically covers a wide range of the sample space and (b) can recover from wrong answers, as participants may provide noisier responses earlier in the test while gaining familiarity.
\end{itemize}
\fi

%\vspace{-0.1in}
\subsection{Training a perceptual metric}
\label{ssec:framework2}
A high quality perceptual distance metric $D$ would provide a small distance $D(x_\mathtt{ref},x_\mathtt{per})$ if human judges feel they are the same recording, and a larger distance if they are judged to be different. Here, we explore four separate strategies to learn such a metric. We then investigate how well each method correlates with human judgments. All models have the same architecture for comparison, described in Section \ref{subsubsec:architecture}.

\noindent {\bf Using a pre-trained network.}
``Off-the-shelf'' deep network embeddings have been used as a metric for training and have been shown to correlate well with human perceptual judgments in the vision setting~\cite{zhang2018unreasonable}, even without being explicitly trained on perceptual human judgments. We first investigate if similar trends hold in the audio setting. We describe the activation of layer $l$ of an L-layer deep network embedding as $F_l(x) \in \mathds{R}^{T_l\times C_l}$, where $T_l$ and $C_l$ are the time resolution and number of channels of the layer, respectively. The distance between two audio recordings can be defined by averaging between the full feature activation stack,
%\vspace{-0.10in}
\begin{equation*}
D(x_\mathtt{ref},x_\mathtt{per})=\sum_l \frac{1}{T_l C_l}  ||\mathcal{F}_l(x_\mathtt{ref})- \mathcal{F}_l(x_\mathtt{per})||_1
\label{eq1}
%\vspace{-0.10in}
\end{equation*}
\par
We train a model (\textit{\textbf{pre}}) on two general audio classification tasks from DCASE 2016~\cite{mesaros2017detection}, namely accoustic scene classification (ASC) and domestic audio tagging (DAT), following the strategy in~\cite{germain2018speech}.

%In the first task, we are provided
%with audio files featuring various scenes and the goal
%is to determine the scene type for each file. In the second task,
%we are given audio files featuring events of interest and the goal %is to determine which events took place in
%each file (with possibly multiple events in one file).

%%%%%%%%%%%%%%%%%%%%%%%%%%%%%%%%%%%%%%%%%%%%%%%%%%%%%%%%%%%%%%%%%%%%%%%%%%%%%%%%
\noindent {\bf Training a model on perceptual data.}
We take the above model $\mathcal{F}$ and add linear weights over it,
%%\vspace{-0.10in}
\begin{equation*}
D(x_\mathtt{ref},x_\mathtt{per})=\sum_l \frac{1}{T_l C_l} ||w_l \odot (\mathcal{F}_l(x_\mathtt{ref})- \mathcal{F}_l(x_\mathtt{per}))||_1
\label{eq2}
%\vspace{-0.05in}
\end{equation*}
where $w_l \in \mathds{R}^C_l$ and $\odot$ is the Hadamard product over channels. The linear weights effectively decide which channels are more or less ``perceptual''. We present three variants. First, we keep the weights of all the layers $\mathcal{F}$ fixed and only train the linear layers. This presents a ``linear calibration'' of an off-the-shelf network, denoted as \textit{\textbf{lin}}. Second, we initialize from a pre-trained classification model (\textit{\textbf{pre}}), and allow all the weights for network ($\mathcal{F}$ and linear layer) to
be fine-tuned, denoted as \textit{\textbf{fin}}. Third, we allow the network $\mathcal{F}$ and the linear layer to both be trained from
% scratch, denoted as
\textit{\textbf{scratch}}.

\noindent {\bf Training objective.}
Our network $\mathcal{F}$ has a small classification network $\mathcal{G}$ at the end, which maps this distance $D(x_\mathtt{ref},x_\mathtt{per})$ to a predicted human judgment $\hat{h}$. We minimize the binary cross-entropy (BCE) between this predicted value and ground truth human judgment~$h$:
% Given human perceptual judgments $h$, we train this entire network using binary cross-entropy loss over these confidence probabilities.
\vspace{-0.20\baselineskip}
\begin{equation*}
    \mathcal{L}(G,D) = \text{BCE}(G(D(x_\mathtt{ref},x_\mathtt{per})), h)
\end{equation*}

%\vspace{-0.1in}
\section{Experimental Setup}
\label{sec:experiments}

We now describe experiments in the framework from Section ~\ref{ssec:framework_main}.
\vspace{-1.5\baselineskip}
\subsection{Perturbation Space}
\label{sec:perts}
We apply the proposed framework to the broad field of speech telecommunication. In this domain, noises like packet losses, jitter, variable delay and other channel noise artifacts like channel noise, and sidetones are common.

Table~\ref{ref-perturbations} lists all perturbations and their ranges we examined. For each listening test set, we select at most one instance from each category and sample a random order to apply these categories, $\eg$ white noise energy from additive, DRR from reverb, MP3 bit rate from compression, equalization (EQ) and pops as the five perturbation values $(v_{1},v_{2},v_{3},v_{4},v_{5})$. We permute the order to simulate different scenarios. For example, (reverb, additive, EQ, compression, pops) simulates telecommunication while (compression, EQ, reverb, pops, additive) simulates playback audio in a room environment.

\def\arraystretch{1.15}
\begin{table}[]
\caption{All examined perturbations and configurations.}
\vspace{-0.1in}
\resizebox{1\columnwidth}{!}{
\begin{tabular}{lll}
\hlineB{4}
{\color[HTML]{333333} \textbf{Category}}               & \textbf{Perturbations}               & \textbf{Intervals/Range}            \\ \hline
{\color[HTML]{333333} }                                & Applause Noise                       &                                     \\
{\color[HTML]{333333} }                                & Pink Noise                           &                                     \\
{\color[HTML]{333333} }                                & Water Drop Noise                     &                                     \\
{\color[HTML]{333333} }                                & White Noise                          &                                     \\
\multirow{-5}{*}{{\color[HTML]{333333} Additive~\cite{piczak2015esc}}}        & Room Noise                           & \multirow{-5}{*}{2 dB to 66 dB SNR} \\ \hline
{\color[HTML]{333333} }                                & Direct to Reverberation Ratio (DRR)  & -27 dB to 65 dB                     \\
\multirow{-2}{*}{{\color[HTML]{333333} Reverb~\cite{traer2016statistics}}}        & Reverberation Time (RT60)            & 0.05 sec to 8 sec                   \\ \hline
{\color[HTML]{333333} }                                & MP3 (bitrate)                        & 8 Kb/sec to 320 Kb/sec              \\
\multirow{-2}{*}{{\color[HTML]{333333} Compression}}   & $\mu$ law compression (re-quantization) & 1 bit to 60 bits                    \\ \hline
{\color[HTML]{333333} Equalization}                    & Frequency bands (cut/boost bands)    & 0 to 1                              \\ \hline
{\color[HTML]{333333} }                                & Pops (\% audio samples)              & 0.01 \% to 10 \%                    \\
{\color[HTML]{333333} }                                & Griffin-lim (iterations)             & 1 to 500                            \\
\multirow{-3}{*}{{\color[HTML]{333333} Miscellaneous}} & Dropouts (\% audio samples)       & 0.01 \% to 20 \%                    \\ \hlineB{4}
\end{tabular}
}

\label{ref-perturbations}
\vspace{-2ex}
\end{table}

% To simulate these issues

% \begin{figure*}[h!]
% \centering
% \vspace{-12ex}
% %\scalebox{<horizontal scale factor>}[<vertical scale factor>]{\includegraphics{<file name>}}
% %\includegraphics[width=50mm,scale=0.5]{method.eps}
% %\scalebox{1}[0.75]{\includegraphics[width=\textwidth]{Interspeech_SE.jpg}}
% \includegraphics[width=\textwidth]{Interspeech_denoising.jpg}
% \vspace{-48ex}
% \caption{Denoising pairwise comparison listening test results. Majority vote says our method is better than the baseline. Each plot lists a specific \textit{ours vs baseline} experiment. Results are divided into tranches based on difficulty. Chance is 50\%.}
% \label{denoising}
% \vspace{-3ex}
% \end{figure*}

%\vspace{-0.1in}
\subsection{Crowdsourcing for Data Collection}
After determining the perturbation space, we crowdsource JND answers on Amazon Mechanical Turk (AMT). We require workers to have above 95\% approval ratings. At the beginning of the Human Intelligence Task (HIT), the subject goes through a volume level calibration test in which loud and soft sounds are played alternatively. 
%This is to make sure that they are not only able to hear big differences but subtle differences as well. 
The participants are then asked not to change the volume in the middle of the HIT. Next, an attention test is presented where the participant is asked to identify a word heard in a long sentence. This removes participants that either do not understand English or were not paying attention.
%because the participants cannot replay the audio once it finishes. 
Upon successfully choosing the right word, the subject goes through two teaching tests, where we train the workers on what kind of differences to look for before we move on to the actual task.
% We observed that most participants were not spending enough time on reading the instructions, and were confused when the actual test started. We then move over to our actual task. 
Each HIT contains 30 pairwise comparisons, 10 each for one randomly chosen reference and direction. 
Out of these 30 comparisons, 6 (20\%) tests are sentinel questions in the form of obvious audio deformations. 
If the participant gets any of the 6 questions wrong, we discard their data. Each audio recording is roughly 2.5 seconds long, and the subjects can replay the files if they choose to. On an average, it takes 7-8 minutes to complete a HIT. 
At the end, we also ask for comments/suggestions/reviews from the participants on their experience in doing this HIT.
% It was really helpful to get feedback to know if participants could understand the question so that we could better phrase the language of our question and add extra teaching tasks. 
We launched 2000 HITs and retained 1812 after validation, collecting about 55k pairs of human subjective judgments.

\vspace{0.5\baselineskip}
\noindent We verify that the resulting dataset has the desired properties:
\begin{enumerate}[leftmargin=0.33cm]
  \item \textbf{Balanced number of same or different answers}: Our active learning strategy predicts the JND of the listener given all their previous answers. JND is a point at which the listener is equally probable to say exactly same or different and so if our model indeed works well, we should observe that there be an almost equal number of ``same'' or ``different'' answers in our dataset. This is precisely what we observe - we observe 25782 ``same'' to 26130 ``different'' answers.
  \item \textbf{Individual consistency checking}: We check consistency between all answers given by a listener and their final predicted JND level; noise levels lower than JND should say ``same`` and higher than JND should say ``different``. Low value of user agreement would mean that the listener answered randomly and/or our method of JND prediction is not accurate, whereas a high value would mean that our model correctly predicts JND and that the users don't answer randomly. The user agreement is around 88.3\% which is high.
  \iffalse
  \item \textbf{Individual consistency checking}: We take all answers given by a listener and compare it with their final predicted JND to see how many answers correlate. Low value of user agreement would mean that the listener answered randomly and/or our method of JND prediction is not accurate, whereas a high value would mean that our model rightly predicts JND and that the users don't answer randomly. The user agreement is around 88.3\% which is high.
  \fi
\end{enumerate}

\subsection{Training and architecture}
\label{subsubsec:architecture}
We use a network inspired by~\cite{germain2018speech} consisting of 14 convolutional layers with 3$\times$1 kernels, batch normalisation and leaky ReLU units, and zero padding to reduce the output dimensions by half after every step. The number of channels double after every 5 layers starting with 32 channels in the first layer. We also use dropout in all convolutional layers. The receptive field of the network is $2^{14}-1$. We train the model using cross-entropy loss using a small classification model that maps distance to predicted human judgment.
%Given two audio clips $x_\mathtt{ref}$ and $x_\mathtt{per}$, we use \eqn{eq1} or \eqn{eq2} depending upon the model to get one singular value of distance estimate between $x_\mathtt{ref}$ and $x_\mathtt{per}$. 

We train this network for 1000 epochs, taking $\approx$ 3 days to complete using 1 GeForce RTX 2080 GPU. As part of online data augmentation to make the model invariant to small delay, we decide randomly if we want to add a 0.25s silence to the audio at the beginning or the end and then present it to the network. This helps providing shift invariance property to the model, to disambiguate that in fact the audio is similar when time shifted.

%\vspace{-0.1in}
\section{Results}

\subsection{Subjective Validation}
We use previously published diverse large-scale third-party studies to verify that our trained metric correlates well on their task. We show results of our models, and compare these with embeddings obtained from self-supervised models (\eg OpenL3~\cite{cramer2019look}) and large-scale pretrained models (\eg VGGish~\cite{hershey2017cnn} trained on Audioset~\cite{gemmeke2017audio}) as well as more conventional objective metrics such as MSE, PESQ~\cite{rix2001perceptual}
%~\footnote{implementation from~\cite{hu2006subjective}} 
and ViSQOL~\cite{beerends2013perceptual}.
%~\footnote{implementation from \url{https://qxlab.ucd.ie/index.php/speech-quality-metrics/}}. 

We compute the correlation between the model's predicted distance with the publicly available MOS scores, using Spearman's Rank order correlation (SC) and Pearson's correlation coefficient (PC). These correlation scores are evaluated per speaker where we average scores for each speaker for each condition.

As an extension, we also check for 2-alternative forced choice test (2AFC) accuracy in which we present one reference recording and two test recordings and ask listeners which one sounds more similar to the reference. Each triplet is evaluated by roughly 10 listeners. 2AFC checks for exact ordering of similarity at per sample basis whereas MOS checks for aggregated ordering, scale and consistency. We choose four distinct classes of available datasets for our analysis:

\iffalse
MOS tests usually have high variance across listeners which makes them too noisy to correlate with. Sometimes, when the data is noisy, there is a big gap between SC and PC scores which is precisely why we also test for 2AFC in which we present one reference audio and two test audio and ask listeners which one sounds more similar to the reference. Each triplet is evaluated by roughly 10 listeners. We choose three distinct classes of publicly available datasets for our analysis:
\fi

\begin{enumerate}[leftmargin=0.33cm]
   \item \textbf{VoCo}~\cite{jin2017voco}:
   consists of MOS tests to verify quality of 6 different word synthesis and insertion algorithms, hence not sample-aligned data.
   
   %\vspace{-0.10in}
   \item \textbf{FFTnet}~\cite{jin2018fftnet}:
   consists of MOS tests for synthetic audio generated by 5 different type of speech generation algorithms. It introduces artifacts specific to SE (speech enhancement), and are not sample-aligned due to phase change. The 2AFC study consists of 2050 triplets of clean reference and noisy test recordings. 
   %\vspace{-0.10in}
   \item \textbf{Bandwidth Expansion}~\cite{feng2019learning}:
   consists of MOS tests for 3 different bandwidth expansion algorithms, aiming at increasing sample rate by filling in the missing high-frequency information. These audio samples consist of very subtle high-frequency differences. The 2AFC study consists of 1020 triplets of clean reference and noisy test recordings.
   
   \item \textbf{Simulated}:
   consists of 1210 triplets of clean reference and noisy test recordings from our perturbation space described in Section~\ref{sec:perts}.
   
\end{enumerate}
\par
The results are displayed in Table~\ref{table_mos}, in which our proposed method ``\textbf{scratch}'' has the best performance overall. We also summarize a few other notable observations, listed below:

\begin{itemize} [leftmargin=0.33cm]
    % noitemsep
    \item Neural-network-based metrics are more robust to non-sample-aligned data - we see that most of our models, including \textit{pre}, perform better than conventional metrics on non-sample-aligned synthetic audio samples in \textit{VoCo} and \textit{FFTnet}.
    
    \item \textit{PESQ} and \textit{VISQOL} have better 2AFC accuracy on \textit{FFTnet} and \textit{Simulated} datasets but lower MOS scores suggesting that these methods preserve ordering but not scale. We also observe that \textit{pre} has higher 2AFC accuracy but slightly lower MOS scores, suggesting that it learns perceptual ordering but not the scale. Interestingly, it performs better than \textit{scratch} on the above two 2AFC datasets, suggesting that training on  (related) classification may produce features that better correlates with ordering (A is closer to C than B is to C). Tuning on JND data could be considered as calibrating on the scale (how much A is close to C).
    
    \iffalse
    correlate with judgment better than what you get by directly training on human perception. \textit{pre} learns large scale features, but not fine grained details which are important in BWE: which can be captured by training on human perception.

    \Pranay{
    
    \item{\textit{PESQ} and \textit{VISQOL} have better 2AFC accuracy across FFTnet and Simulated datasets but lower MOS scores, suggesting that the. Also about the }
    
    \textit{pre} performs well, suggesting that the model learns features that correlate perceptually with human judgments. Interestingly, it performs better than \textit{scratch} on 2 2AFC datasets: suggesting that training on  (related) classification tasks can learn features which correlate with judgment better than what you get by directly training on human perception. \textit{pre} learns large scale features, but not fine grained details which are important in BWE: which can be captured by training on human perception.}
    \fi
    
    \item Though the \textit{pre} model is robust to non-sample-aligned data, it has problems on revealing high frequency subtle difference. It is likely because this model is trained on a task ( sound classification) that does not rely on these frequency bands. On the contrary \textit{VGGish} and \textit{OpenL3} perform relatively well as they are trained on much larger-scaled tasks (AudioSet~\cite{gemmeke2017audio}) and thus are more reliant on high frequency perceptual features. However, they perform worse on the first two tasks, which may be because they are not trained directly on speech. 
    
    \item Conventional metrics such as PESQ and ViSQOL perform better on \textit{VoCo} and \textit{FFTnet} than \textit{BWE}, indicating they are less correlated to human perception when measuring subtle differences.
    
    \iffalse
    \Pranay{\textit{PESQ} and \textit{VISQOL} perform better than our metric on two datasets: both datasets considering large scale distances. But it fails on BWE suggesting that they fail to take fine grained details into consideration. Moreover, by design, our metric is optimised for JND-like small scale distances which is why it performs extremely well on BWE. The next step would be combine the large and small scale models into one.}
    %%\vspace{-0.10in}
    \fi
    %This is because it is trained in a self-supervised manner compared to VGGish which is trained under supervised setting.
    %\vspace{-0.10in}
    \item Methods relying on spectrogram differences (\eg \textit{VGGish}, \textit{OpenL3}, \textit{MSE}) correlate poorly with MOS. Although spectrogram is relatively robust to small shifts, the change in phase can destabilize the amplitude~\cite{purwins2019deep}, causing random variations. When two signals are very similar, this variation becomes dominant causing spectrogram differences to fluctuate and hence, decorrelate with human perception as we see in \textit{VoCo} and \textit{FFTnet} cases. 
    % This shows that (for some cases) loosing out on phase information leads to a decrease in audio quality~\cite{purwins2019deep}, making it one of the disadvantages of using magnitude spectrogram as an input.
    %\vspace{-0.10in}
    \item \textit{OpenL3} performs better than \textit{VGGish} across tasks. It may be because \textit{OpenL3} was trained using self-supervision, mapping both visual and audio onto the same embedded space, which preserves more information than classification.
\end{itemize}

\begin{table}
\centering
\resizebox{\columnwidth}{!}{
 \begin{tabular}{l l c c c c c c c c c}
 \toprule
 \multirow{2}{*}{\bf Type} & \multirow{2}{*}{\bf Name} & \multicolumn{2}{c}{\bf VoCo~\cite{jin2017voco}} & \multicolumn{3}{c}{\bf FFTnet~\cite{jin2018fftnet}}& \multicolumn{3}{c}{\bf BWE~\cite{feng2019learning}}& 
 \multicolumn{1}{c}{\bf Simulated} \\
 \cmidrule(lr){3-4} \cmidrule(lr){5-7} \cmidrule(lr){8-10} \cmidrule(lr){11-11} 
 & &\bf SC &\bf PC & \bf SC & \bf PC & \bf 2AFC & \bf SC & \bf PC & \bf 2AFC & \bf 2AFC \\
 \cmidrule(lr){1-11}
\multirow{4}{*}{\bf Ours}
 & {\bf Pre} & 0.60 & 0.90  & 0.37 & 0.32 & 77.3 & 0.00 & 0.23 & 70.5 & 83.9 \\
 & {\bf Lin} & 0.30 & 0.45  & 0.40 & 0.29 & 77 & 0.48 & 0.36 & 76.6 & 83.6   \\
 & {\bf Fin} & 0.46 & 0.71 & 0.45 & 0.30  & 73.5 &  0.50 & {\bf 0.66} & 86 & 80.3  \\
 & {\bf Scratch} & {\bf 0.71} & {\bf 0.94}  & {\bf 0.63} & \bf 0.59 & 70 & {\bf 0.61} & 0.47 & \bf 87.68 & 71.78 \\ 
 \cdashline{1-11}
 \multirow{2}{*}{\bf Self-sup}
  & {\bf VGGish} & 0.10 & 0.23  & -0.41 & -0.44  & 63 & 0.51 & 0.50 & 52.3 & 76.2 \\
 & {\bf OpenL3} & 0.27 & 0.36 & 0.12 & 0.17 & 65.2 &  0.53 & 0.53 & 61.1 & 73.5  \\
  \cdashline{1-11}
  \multirow{3}{*}{\bf Conv}
  & {\bf MSE} & 0.18 & 0.80 & 0.18 & 0.15 & 66 & 0.00 & 0.26 & 49 & 43  \\
 & {\bf PESQ} & 0.43 & 0.85 & 0.49 & 0.56 & \bf 88.57 & 0.21 & 0.18 & 38.1 & \bf 86.1 \\
 & {\bf ViSQOL} & 0.50 & 0.75 & 0.02 & 0.35 & 79 & 0.13 & 0.09 & 44.4 & 84.2 \\
 \bottomrule
\end{tabular}
}
\caption{Spearman (SC), Pearson (PC) and 2AFC accuracy. Models include: ours, (self)-supervised embeddings, and conventional metrics. $\uparrow$ is better.}
\label{table_mos}
\vspace{-5ex}
%\vspace{-0.40in}
%%%\vspace{-.05in}
\end{table}

\begin{figure}[h!]
\centering
\includegraphics[width=\columnwidth]{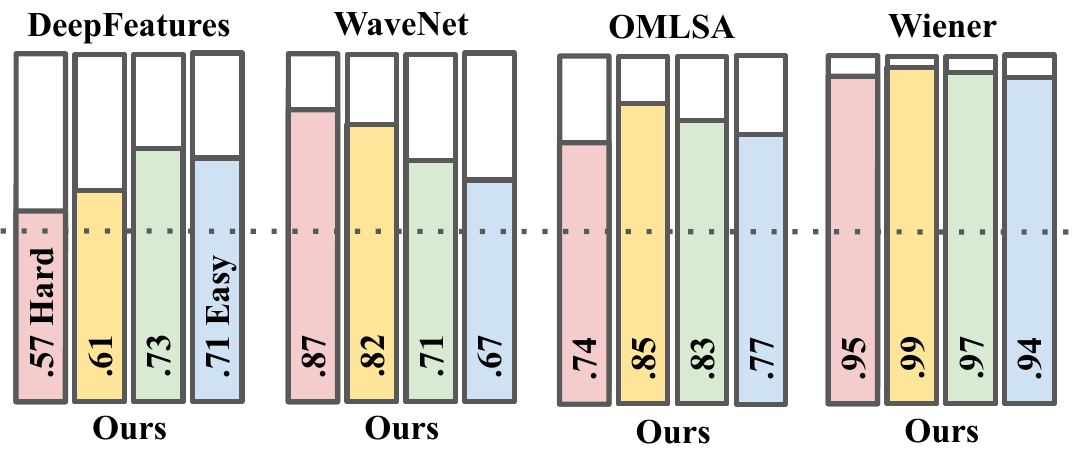}
\vspace{-4ex}
\caption{Pairwise comparison of denoising methods. Our method (below) is preferred over each baseline (above). Results are divided into tranches based on difficulty from pink=hard to blue=easy. Chance is 50\% (dotted line). $\uparrow$ is better for ours.}
\label{denoising}
\vspace{-3ex}
\end{figure}

\subsection{Speech enhancement using trained loss}
We show utility of our trained metric as a loss function for Speech Enhancement (SE) task. We use the dataset available in~\cite{valentini2016speech}, consisting of around 11,572 utterances for training and 824 files for validation. This dataset consists of 28 speakers equally split between male and female speakers containing 10 unique background types across 4 different SNR's. Our denoising network (separate from the learned loss network) consists of 16 layers of fully convolutional context-aggregation network as in~\cite{germain2018speech}. We keep the size of the SE model the same for fair comparison. We compute losses from all 14 layers of our trained loss function and add them to get the total loss. We use Adam Optimiser~\cite{kingma2014adam} with a learning rate of $10^{-4}$, and train for 400 epochs.

We compare our SE model with the state-of-the-art method based on deep feature loss~\cite{germain2018speech} and a few other notable baselines including Wavenet~\cite{rethage2018wavenet}, OMLSA~\cite{cohen2001speech} and Wiener Filter~\cite{hu2006subjective}. We randomly select 600 noisy recordings from the validation set of~\cite{valentini2016speech} and enhance them using the above algorithms. We perform A/B preference tests on AMT, consisting of \textit{Ours vs baseline} pairwise comparisons.
Each pair is rated by 15 different turkers and then majority voted to see which method performs better. To analyse these results, we divide our results into 4 groups: from having highest input noise (Hard) to lowest input noise (Easy). Results are shown in Figure~\ref{denoising}. All results are statistically significant with p $<$ $10^{-4}$.

We observe that our model is preferred on all levels of noises across all baseline methods. Specifically looking at \textit{ours vs deepfeatures}, we observe that our model does best for high SNR recordings, where the degradation caused strong background noise is less noticeable. This highlights the usefulness of training our perceptual loss on JND data. Given that our loss function is trained on JND data, it is able to better correlate with local, subtle differences than other loss functions.

\section{Conclusion and Future Work}
We propose a framework to collect human ``just noticeable difference'' judgments on audio signals. Directly \textit{learning} a perceptual metric from our data produces a metric that correlates better with MOS tests than traditional metrics, such as PESQ~\cite{beerends2013perceptual}. We also showed that a model pre-trained on classification tasks has the correct order but incorrect perceptual scale which can be calibrated by training on our data. Furthermore, we show that the metric can be directly \textit{optimized} as a loss function, in the task of speech enhancement. A similar story has emerged in the computer vision literature, where trained networks have been shown to both correlate well with human perceptual judgments~\cite{zhang2018unreasonable} and serve well as an optimization objective~\cite{johnson2016perceptual}, compared to traditional metrics such as SSIM~\cite{wang2004image}. 

We would like to extend this dataset in the future to explore a broader range of types of perturbations, and do so with a greater density of samples at a wider range of intensities. We would also like to include content beyond speech, particularly music. Such data could be leveraged to study more broadly the manifold of audio perception, and enable a broader set of applications.

\clearpage
\newpage
\bibliographystyle{IEEEtran}

\bibliography{mybib}

\begin{thebibliography}{10}
\providecommand{\url}[1]{#1}
\csname url@samestyle\endcsname
\providecommand{\newblock}{\relax}
\providecommand{\bibinfo}[2]{#2}
\providecommand{\BIBentrySTDinterwordspacing}{\spaceskip=0pt\relax}
\providecommand{\BIBentryALTinterwordstretchfactor}{4}
\providecommand{\BIBentryALTinterwordspacing}{\spaceskip=\fontdimen2\font plus
\BIBentryALTinterwordstretchfactor\fontdimen3\font minus
  \fontdimen4\font\relax}
\providecommand{\BIBforeignlanguage}[2]{{%
\expandafter\ifx\csname l@#1\endcsname\relax
\typeout{** WARNING: IEEEtran.bst: No hyphenation pattern has been}%
\typeout{** loaded for the language `#1'. Using the pattern for}%
\typeout{** the default language instead.}%
\else
\language=\csname l@#1\endcsname
\fi
#2}}
\providecommand{\BIBdecl}{\relax}
\BIBdecl

\bibitem{oord2016wavenet}
A.~v.~d. Oord, S.~Dieleman \emph{et~al.}, ``Wavenet: A generative model for raw
  audio,'' \emph{arXiv preprint arXiv:1609.03499}, 2016.

\bibitem{rix2001perceptual}
A.~W. Rix and J.~G. Beerends~et al., ``Perceptual evaluation of speech quality
  (pesq)-a new method for speech quality assessment of telephone networks and
  codecs,'' in \emph{2001 IEEE International Conference on Acoustics, Speech,
  and Signal Processing. Proceedings (Cat. No. 01CH37221)}, vol.~2.\hskip 1em
  plus 0.5em minus 0.4em\relax IEEE, 2001, pp. 749--752.

\bibitem{beerends2013perceptual}
J.~G. Beerends, C.~Schmidmer \emph{et~al.}, ``Perceptual objective listening
  quality assessment (polqa), the third generation itu-t standard for
  end-to-end speech quality measurement part i—temporal alignment,''
  \emph{Journal of the Audio Engineering Society}, vol.~61, no.~6, pp.
  366--384, 2013.

\bibitem{hines2015visqol}
A.~Hines, J.~Skoglund, A.~C. Kokaram, and N.~Harte, ``Visqol: an objective
  speech quality model,'' \emph{EURASIP Journal on Audio, Speech, and Music
  Processing}, vol. 2015, no.~1, pp. 1--18, 2015.

\bibitem{hines2013robustness}
A.~Hines, J.~Skoglund, A.~Kokaram, and N.~Harte, ``Robustness of speech quality
  metrics to background noise and network degradations: Comparing visqol, pesq
  and polqa,'' in \emph{2013 IEEE International Conference on Acoustics, Speech
  and Signal Processing}.\hskip 1em plus 0.5em minus 0.4em\relax IEEE, 2013,
  pp. 3697--3701.

\bibitem{manjunath2009limitations}
T.~Manjunath, ``Limitations of perceptual evaluation of speech quality on voip
  systems,'' in \emph{2009 IEEE International Symposium on Broadband Multimedia
  Systems and Broadcasting}.\hskip 1em plus 0.5em minus 0.4em\relax IEEE, 2009,
  pp. 1--6.

\bibitem{zhang2018training}
H.~Zhang, X.~Zhang, and G.~Gao, ``Training supervised speech separation system
  to improve stoi and pesq directly,'' in \emph{2018 IEEE International
  Conference on Acoustics, Speech and Signal Processing (ICASSP)}.\hskip 1em
  plus 0.5em minus 0.4em\relax IEEE, 2018, pp. 5374--5378.

\bibitem{fu2019learning}
S.-W. Fu, C.-F. Liao, and Y.~Tsao, ``Learning with learned loss function:
  Speech enhancement with quality-net to improve perceptual evaluation of
  speech quality,'' \emph{IEEE Signal Processing Letters}, 2019.

\bibitem{pascual2017segan}
S.~Pascual, A.~Bonafonte, and J.~Serra, ``Segan: Speech enhancement generative
  adversarial network,'' \emph{arXiv preprint arXiv:1703.09452}, 2017.

\bibitem{donahue2018adversarial}
C.~Donahue, J.~McAuley, and M.~Puckette, ``Adversarial audio synthesis,''
  \emph{arXiv preprint arXiv:1802.04208}, 2018.

\bibitem{stoller2018adversarial}
D.~Stoller, S.~Ewert, and S.~Dixon, ``Adversarial semi-supervised audio source
  separation applied to singing voice extraction,'' in \emph{2018 IEEE
  International Conference on Acoustics, Speech and Signal Processing
  (ICASSP)}.\hskip 1em plus 0.5em minus 0.4em\relax IEEE, 2018, pp. 2391--2395.

\bibitem{gatys2015neural}
L.~A. Gatys, A.~S. Ecker, and M.~Bethge, ``A neural algorithm of artistic
  style,'' \emph{arXiv preprint arXiv:1508.06576}, 2015.

\bibitem{zhang2018unreasonable}
R.~Zhang, P.~Isola, A.~A. Efros, E.~Shechtman, and O.~Wang, ``The unreasonable
  effectiveness of deep features as a perceptual metric,'' in \emph{Proceedings
  of the IEEE Conference on Computer Vision and Pattern Recognition}, 2018, pp.
  586--595.

\bibitem{germain2018speech}
F.~G. Germain, Q.~Chen, and V.~Koltun, ``Speech denoising with deep feature
  losses,'' \emph{arXiv preprint arXiv:1806.10522}, 2018.

\bibitem{ananthabhotla2019towards}
I.~Ananthabhotla, S.~Ewert, and J.~A. Paradiso, ``Towards a perceptual loss:
  Using a neural network codec approximation as a loss for generative audio
  models,'' in \emph{Proceedings of the 27th ACM International Conference on
  Multimedia}, 2019, pp. 1518--1525.

\bibitem{kilgour2018fr}
K.~Kilgour, M.~Zuluaga, D.~Roblek, and M.~Sharifi, ``Fr$\backslash$'echet audio
  distance: A metric for evaluating music enhancement algorithms,'' \emph{arXiv
  preprint arXiv:1812.08466}, 2018.

\bibitem{avila2019intrusive}
A.~R. Avila, J.~Alam, D.~O’Shaughnessy, and T.~H. Falk, ``Intrusive quality
  measurement of noisy and enhanced speech based on i-vector similarity,'' in
  \emph{2019 Eleventh International Conference on Quality of Multimedia
  Experience (QoMEX)}.\hskip 1em plus 0.5em minus 0.4em\relax IEEE, 2019, pp.
  1--5.

\bibitem{cartwright2016fast}
M.~Cartwright, B.~Pardo, G.~J. Mysore, and M.~Hoffman, ``Fast and easy
  crowdsourced perceptual audio evaluation,'' in \emph{2016 IEEE International
  Conference on Acoustics, Speech and Signal Processing (ICASSP)}.\hskip 1em
  plus 0.5em minus 0.4em\relax IEEE, 2016, pp. 619--623.

\bibitem{cartwright2018crowdsourced}
M.~Cartwright, B.~Pardo, and G.~J. Mysore, ``Crowdsourced pairwise-comparison
  for source separation evaluation,'' in \emph{2018 IEEE International
  Conference on Acoustics, Speech and Signal Processing (ICASSP)}.\hskip 1em
  plus 0.5em minus 0.4em\relax IEEE, 2018, pp. 606--610.

\bibitem{mcshefferty2015just}
D.~McShefferty, W.~M. Whitmer, and M.~A. Akeroyd, ``The just-noticeable
  difference in speech-to-noise ratio,'' \emph{Trends in hearing}, vol.~19, p.
  2331216515572316, 2015.

\bibitem{mesaros2017detection}
A.~Mesaros, T.~Heittola, E.~Benetos, P.~Foster, M.~Lagrange, T.~Virtanen, and
  M.~D. Plumbley, ``Detection and classification of acoustic scenes and events:
  Outcome of the dcase 2016 challenge,'' \emph{IEEE/ACM Transactions on Audio,
  Speech, and Language Processing}, vol.~26, no.~2, pp. 379--393, 2017.

\bibitem{piczak2015esc}
K.~J. Piczak, ``Esc: Dataset for environmental sound classification,'' in
  \emph{Proceedings of the 23rd ACM international conference on Multimedia},
  2015, pp. 1015--1018.

\bibitem{traer2016statistics}
J.~Traer and J.~H. McDermott, ``Statistics of natural reverberation enable
  perceptual separation of sound and space,'' \emph{Proceedings of the National
  Academy of Sciences}, vol. 113, no.~48, pp. E7856--E7865, 2016.

\bibitem{cramer2019look}
J.~Cramer, H.-H. Wu, J.~Salamon, and J.~P. Bello, ``Look, listen, and learn
  more: Design choices for deep audio embeddings,'' in \emph{ICASSP 2019-2019
  IEEE International Conference on Acoustics, Speech and Signal Processing
  (ICASSP)}.\hskip 1em plus 0.5em minus 0.4em\relax IEEE, 2019, pp. 3852--3856.

\bibitem{hershey2017cnn}
S.~Hershey, S.~Chaudhuri, D.~P. Ellis \emph{et~al.}, ``Cnn architectures for
  large-scale audio classification,'' in \emph{2017 ieee international
  conference on acoustics, speech and signal processing (icassp)}.\hskip 1em
  plus 0.5em minus 0.4em\relax IEEE, 2017, pp. 131--135.

\bibitem{gemmeke2017audio}
J.~F. Gemmeke, D.~P. Ellis \emph{et~al.}, ``Audio set: An ontology and
  human-labeled dataset for audio events,'' in \emph{2017 IEEE International
  Conference on Acoustics, Speech and Signal Processing (ICASSP)}.\hskip 1em
  plus 0.5em minus 0.4em\relax IEEE, 2017, pp. 776--780.

\bibitem{jin2017voco}
Z.~Jin, G.~J. Mysore, S.~Diverdi, J.~Lu, and A.~Finkelstein, ``Voco: text-based
  insertion and replacement in audio narration,'' \emph{ACM Transactions on
  Graphics (TOG)}, vol.~36, no.~4, pp. 1--13, 2017.

\bibitem{jin2018fftnet}
Z.~Jin, A.~Finkelstein, G.~J. Mysore, and J.~Lu, ``Fftnet: A real-time
  speaker-dependent neural vocoder,'' in \emph{2018 IEEE International
  Conference on Acoustics, Speech and Signal Processing (ICASSP)}.\hskip 1em
  plus 0.5em minus 0.4em\relax IEEE, 2018, pp. 2251--2255.

\bibitem{feng2019learning}
B.~Feng, Z.~Jin, J.~Su, and A.~Finkelstein, ``Learning bandwidth expansion
  using perceptually-motivated loss,'' in \emph{ICASSP}, May 2019.

\bibitem{purwins2019deep}
H.~Purwins, B.~Li, T.~Virtanen, J.~Schl{\"u}ter, S.-Y. Chang, and T.~Sainath,
  ``Deep learning for audio signal processing,'' \emph{IEEE Journal of Selected
  Topics in Signal Processing}, vol.~13, no.~2, pp. 206--219, 2019.

\bibitem{valentini2016speech}
C.~Valentini-Botinhao, X.~Wang, S.~Takaki, and J.~Yamagishi, ``Speech
  enhancement for a noise-robust text-to-speech synthesis system using deep
  recurrent neural networks.'' in \emph{Interspeech}, 2016, pp. 352--356.

\bibitem{kingma2014adam}
D.~P. Kingma and J.~Ba, ``Adam: A method for stochastic optimization,''
  \emph{arXiv preprint arXiv:1412.6980}, 2014.

\bibitem{rethage2018wavenet}
D.~Rethage, J.~Pons, and X.~Serra, ``A wavenet for speech denoising,'' in
  \emph{2018 IEEE International Conference on Acoustics, Speech and Signal
  Processing (ICASSP)}.\hskip 1em plus 0.5em minus 0.4em\relax IEEE, 2018, pp.
  5069--5073.

\bibitem{cohen2001speech}
I.~Cohen and B.~Berdugo, ``Speech enhancement for non-stationary noise
  environments,'' \emph{Signal processing}, vol.~81, no.~11, pp. 2403--2418,
  2001.

\bibitem{hu2006subjective}
Y.~Hu and P.~C. Loizou, ``Subjective comparison of speech enhancement
  algorithms,'' in \emph{2006 IEEE International Conference on Acoustics Speech
  and Signal Processing Proceedings}, vol.~1.\hskip 1em plus 0.5em minus
  0.4em\relax IEEE, 2006, pp. I--I.

\bibitem{johnson2016perceptual}
J.~Johnson, A.~Alahi, and L.~Fei-Fei, ``Perceptual losses for real-time style
  transfer and super-resolution,'' in \emph{European conference on computer
  vision}.\hskip 1em plus 0.5em minus 0.4em\relax Springer, 2016, pp. 694--711.

\bibitem{wang2004image}
Z.~Wang, A.~C. Bovik \emph{et~al.}, ``Image quality assessment: from error
  visibility to structural similarity,'' \emph{IEEE transactions on image
  processing}, vol.~13, no.~4, pp. 600--612, 2004.

\end{thebibliography}

\clearpage
\newpage

\section{Supplementary Material}
\beginsupplement

\subsection{Details about framework}
Refer to section \ref{ssec:framework1}. Note that our goal is to measure JND which is to look for $\rho_{jnd}$ that makes the difference between $x_\mathtt{ref}$ and $x_\mathtt{per}$ just noticeable. Additionally, we put priors on $\mu$ and $\sigma$ to make the first several tests less susceptible to human error. This has several advantages: 
\begin{enumerate}
  \item \textbf{Information Maximization}: one good way to achieve maximum information gain is to ask questions around JND, which is where the answers are the least obvious and most challenging. Also, this strategy has the advantage of inherently creating a ``balanced`` dataset~\footnote{ N. Roy and A. McCallum - Toward optimal active learning through monte carlo estimation of error reduction - ICML 2001} where you have an almost equal number of ``same`` or ``different`` answers.
  \item \textbf{Extra added bias}: We also encourage an equal chance of saying \textit{same} or \textit{different} by using $v_{\mathtt{jnd}}^*=\mu+q\sigma$, where $q > 0$ when we have collected more ``same'' than ``different'', and vice versa. This is done so that the participant is likely to break the trend of giving same answers. If the same trend still continues, we discard that participants' data as that are not paying attention.
  
  \item \textbf{Additional Priors}: Our model also starts out with a prior that focuses on exploration early on in the test. As more data is acquired, the model becomes more confident and the prior is deemphasized. This procedure (a) stochastically covers a wide range of the sample space and (b) can recover from wrong answers, as participants may provide noisier responses earlier in the test while gaining familiarity.
\end{enumerate}

\subsection{Details about perturbation space}
Refer to Section~\ref{sec:perts}. Table~\ref{ref-perturbations} lists all perturbations we examined with their range. We choose these perturbations to simulate the commonly found artifacts in the broad field of speech telecommunication.
We divide our perturbation set into five categories:
\begin{itemize}[leftmargin=0.33cm]
   
    \item \textit{Additive perturbations}: include noises like applause, pink noise, water drop noise, white noise and room noise taken from ESC50~\cite{piczak2015esc}.
    
    \item \textit{Reverb perturbations}: we use a dataset of real impulse responses (IR)~\cite{traer2016statistics} and approximately modify  the Direct-to-Reverberant Ratio (DRR) and Reverberation Time (RT60) of the sampled IR by multiplying it with a constant after the first direct response and time stretching respectively.
    
    \item \textit{Compression perturbations}: we consider $\mu$-law encoding, where we change the number of bits to encode the audio and MP3 compression, and vary the bit-rate.
    
    \item \textit{Equalization(EQ) distortion}: we consider three frequency bands where we cut or boost a range of frequencies.
    
    \item \textit{Miscellaneous perturbations}: we also consider three miscellaneous perturbations like \textit{pops} (simulate the popping artifacts on telephones), \textit{griffin-lim} (regenerate phase) and \textit{dropouts} (silent random parts of audio).
    
\end{itemize}

\newpage

\subsection{Training Details}
Here, we provide some additional details on model training for our networks trained on distortions. We train with $1000$ epochs at a learning rate of $10^{-4}$ with batch size $16$. We enforce non-negative weightings on the linear layer $w$, since larger distances in
a certain feature should not result in two recordings becoming
closer in the distance metric. This is done by projecting the
weights into the constraint set at every iteration. In other
words, we check for any negative weights, and force them
to be 0. The project was implemented using Tensorflow~\footnote{M. Abadi and P. Barham - Tensorflow: A system for large-scale machine learning - OSDI 2016}

\end{document}